\documentclass[prl,aps]{revtex4}
\usepackage{epsfig}
\begin{document}

\title{Agterberg and Dodgson Reply}
\author{D.F. Agterberg$^1$ and Matthew J.W. Dodgson$^{2,3}$}

\address{$^1$Department of Physics, University of Wisconsin-Milwaukee, Milwaukee, WI 53201\\
$^2$Theory of Condensed Matter Group, Cavendish Laboratory, Cambridge, CB3 0HE, United Kingdom\\
$^3$Institut de Physique, Universit\'e de Neuch\^atel, 2000
Neuch\^atel, Switzerland}

\begin{abstract}
A reply to the Comment by Mineev and Champel.
\end{abstract}
\maketitle

In their Comment \cite{min02}, Mineev and Champel have argued that
our results are incorrect because of the simultaneous neglect of
both non-linear and non-local terms in our theory.  We disagree
with this statement and point out that while Ref.~\cite{min02}
argue that these terms should be included, they do not give a
convincing physical argument that these terms are important at low
fields. In the following, we explicitly show that non-linear and
non-local terms can be safely ignored {\it near} $H_{c1}$ where
our results are valid. Furthermore, our results agree with the
assertion of Ref.~\cite{min02} that there is no true $A$ to $B$
phase transition in the vortex phase. In Ref.~\cite{agt02}, we
refer to the diverging correlation length that describes the $A$
to $B$ transition in the {\it Meissner} phase.


To justify our claims, we give the missing steps between Eq.~2 and
Eq.~3 of Ref.~\cite{agt02}. The expression used for $\Psi_-$ in
the $A$ phase was
\begin{equation}
(\xi_{A,-}^{-2}-{\bf D}^2)\Psi_-({\bf
r})=\frac{\tilde{\kappa}}{\kappa}(D_xD_y+D_yD_x)\Psi_+({\bf r})
\label{eq1}
\end{equation}
where $D_i=\partial_i+i2\pi A_i/\Phi_0$.  Eq.~3 in
Ref.~\cite{min02} is found by setting the $-{\bf D}^2$ operator to
zero. This operator cuts off the divergence that Ref.~\cite{min02}
point out in their Eq.~3. We work with
$\tilde{\kappa}/\kappa\ll 1$
and within a London approach.
Taking $\Psi_+({\bf r})=|\Psi_+|e^{i\phi}$,
$\Psi_-=e^{i\phi}\tilde{\Psi}_-$, and defining the superfluid
velocity as
${\bf v}=\nabla \phi+2\pi{\bf A}/\Phi_0$
we get
\begin{equation} (\xi_-^{-2}+{\bf
v}^2-2i{\bf v}\cdot{\nabla}-\nabla^2)\tilde{\Psi}_-({\bf
r})=|\Psi_+|\frac{\tilde{\kappa}}{\kappa}(i\partial_x
v_y+i\partial_y v_x -2v_xv_y) \label{eq2}.
\end{equation}
This last equation shows that $\tilde{\Psi}_-$ will be
proportional to $v$
(for small $v$).
Consequently, for large vortex separations, the non-linear term in
$v$ on the right hand side and the terms with $v$ on the left hand
side of Eq.~\ref{eq2} can be ignored
($v\propto e^{-r/\lambda}/\sqrt{r\lambda}$
and
the derivatives give a factor
$1/\lambda\gg v$
for large $r$). After
Fourier
transforming
and using
the Maxwell relation
${\bf v}=\frac{2\pi\lambda^2}{\Phi_0}\nabla\times{\bf B}$,
Eq.~\ref{eq2}
becomes
\begin{equation}
\tilde{\Psi}_-({\bf
q})=|\Psi_+|\frac{\tilde{\kappa}}{\kappa}\frac{q_x^2-q_y^2}{\xi_{A,-}^{-2}+{\bf
q}^2}\frac{2 \pi \lambda^2}{\Phi_0}B({\bf q}). \label{eq3}
\end{equation}
Using this expression in the free energy and minimizing with
respect to $B({\bf q})$ gives the novel London equation that forms
the basis of our results
\begin{equation}
\Big [1 + \lambda^2 {\bf q}^2 +
\frac{\lambda^2\tilde{\kappa}^2}{\kappa^2}\frac {(q_x^2 -
q_y^2)^2}{\xi_{A,-}^{-2} + {\bf q}^2}\Big ] B({\bf q}) = 0.
\end{equation}
This gives a 4-fold symmetry to the structure of a flux line out
to the distance $\xi_{A,-}$, which diverges at the A$\rightarrow$B
transition temperature.

Using Eq.~\ref{eq3},
 $|\Psi_-({\bf r})|\lesssim\frac{\tilde{\kappa}}{\kappa}|\Psi_+|\frac{2\pi \lambda^2
B({\bf r})}{\Phi_0}$ (take $q_x^2-q_y^2 \rightarrow {\bf q}^2$ in
Eq.~\ref{eq3}). This, with $\beta_1 |\psi_+|^2/\kappa = 1/\xi_+^2$
and $2\pi \lambda^2 B({\bf r})/\Phi_0=\lambda^2/d^2$, we find:
\begin{equation}
\frac{\langle \beta_1 |\Psi_-|^4 \rangle}{\langle \kappa |{\bf D}
\Psi_-|^2\rangle}
\lesssim
\frac{\tilde{\kappa}^2}{\kappa^2}\frac{\lambda^2}{\xi_{A,+}^2}\frac{\lambda^4}{d^4}
\end{equation}
where $d$ is the distance between vortices. Ref.~\cite{min02} uses
this result as their Eq.~5. $H_{c1}$ marks a second order phase
transition between the Meisner and the vortex phase. At this
transition $B=0$, which implies that as $H\rightarrow H_{c1}$,
$d\rightarrow \infty$. Consequently, sufficiently near $H_{c1}$,
the non-linear term can be safely neglected. Mineev and Champel
argue that $d$ is cutoff when $d\approx
\lambda/\sqrt{\ln(\lambda/\xi_{A,+})}$. Presumably, this accounts
for the very small range of $H$ over which $B$ goes from $H_{c1}$
to zero (or $d$ goes from $\lambda$ to $\infty$). Nevertheless,
this range is experimentally accessible, and furthermore, for flat
plate-like samples the demagnetization factors force the applied
field $H_{app}\propto B$, thus making the large $d$ limit even
more accessible. It is also argued that when Eq.~6 in
Ref.\cite{min02} is not satisfied, then non-local corrections must
be included (these are terms that are $O({\bf q}^4)$ or larger in
Eq.~4). To counter this, we point out that Eq.~4 implies new
physics at low fields (for small $q$ and large $d$) while
non-local terms give new physics at high fields (for large $q$ and
small $d$). In fact, we find non-local terms are not important
when
$\tilde{\kappa}/\kappa>\frac{\xi_{A,+}}{\lambda}\frac{\lambda}{d}$.
Consequently, non-local terms also become negligible in the large
$d$ limit.

DFA was supported by an award from Research Corporation. MJWD was
supported by an EPSRC Advanced Fellowship AF/99/0725.

\end{document}